**PHOTONICS Research**

# Time shifting deviation method enhanced laser interferometry: ultrahigh precision localizing of traffic vibration using a urban fiber link

GUAN WANG,[1,2,†] ZHONGWANG PANG,[1,2,†] BOHAN ZHANG,[1] FANGMIN WANG,[1,2] YUFENG CHEN,[1,2] HONGFEI DAI,[1,2] BO WANG,[1,2,*] AND LIJUN WANG[1,2]

[1]State Key Laboratory of Precision Measurement Technology and Instruments, Department of Precision Instrument, Tsinghua University, Beijing 100084, China
[2]Key Laboratory of Photonic Control Technology (Tsinghua University), Ministry of Education, Beijing 100084, China
*Corresponding author: bo.wang@tsinghua.edu.cn

Using a fiber network as a huge sensing system will enrich monitoring methods of public infrastructures and geological disasters. With the traditional cross-correlation method, a laser interferometer has been used to detect and localize the vibration event. However, the random error induced by the cross-correlation method limits the localization accuracy and makes it not suitable for ultrahigh precision localizing applications. We propose a novel time shifting deviation (TSDEV) method, which has advantages over the cross-correlation method in practicability and localization accuracy. Three experiments are carried out to demonstrate the novelty of the TSDEV method. In a lab test, vibration localization accuracy of ∼2.5 m is realized. In field tests, TSDEV method enhanced interferometry is applied to monitor the urban fiber link. Traffic vibration events on the campus road and Beijing ring road have been precisely localized and analyzed, respectively. The proposed technique will extend the function of the existing urban fiber network, and better serve the future smart city.    © 2022 Chinese Laser Press

https://doi.org/10.1364/PRJ.443019

## 1. INTRODUCTION

As one of the largest infrastructures, the optical fiber network has shown potential in more and more applications. Beside the basic data transmission function, it can be utilized as medium for time-frequency dissemination in the metrological field [1–12] and can also serve as the sensing medium to detect vibration events in the geodesy and earth observation fields [13–19]. Considering the widely distributed urban fiber links, it is likely to monitor public infrastructures in urban area, such as traffic vibrations, which will enrich the sensing method of a smart city.

In our previous study, we constructed a fiber-based time-frequency network in Beijing [Fig. 1(a)]. Through actively compensating the phase fluctuation, several urban fiber links were employed to realize a real-time frequency comparison network, which included five hydrogen masers from four institutions [20,21]. On the other hand, this fiber network has the potential to serve as a huge interferometer to detect vibrations along it. Generally, fiber vibration sensing techniques can be sorted as forward-transmission [22–26] and backscattering [27–33] schemes. The backscattering scheme, such as the distributed acoustic sensing, is a mature technique. However, great optical loss limits its usage range. Furthermore, due to the backscattering detection and the high-power injection pulses, it normally needs to occupy a dark fiber alone. In our case, we choose the forward-transmission scheme, which can be integrated with fiber communication function [15]. However, the main problem of this scheme is how to precisely localize vibration events. This problem limits its application area compared with the backscattering scheme.

Reference [15] adopts the cross-correlation method to localize earthquakes and realizes a localization accuracy of kilometer magnitude in the lab test. It is precise enough for earthquake detection but cannot meet the ultrahigh precision requirement for the scenario of urban vibration detection. Reference [26] uses a frequency-shifted optical delay line and realizes vibration localization on 615 km in-lab link. For vibration signals of 250 Hz, the localization accuracy is 553 m. Its accuracy cannot meet the requirement of ultrahigh precision localizing cases, and the frequency range is also not suitable for vibration events with low frequency (earthquake, traffic, etc.). Actually, cross correlation is a powerful method when evaluating the similarity of two time-infinite signals or time-finite period signals but will bring nonnegligible random error when dealing with aperiodic vibration signals [34]. This random error depends on the cutoff

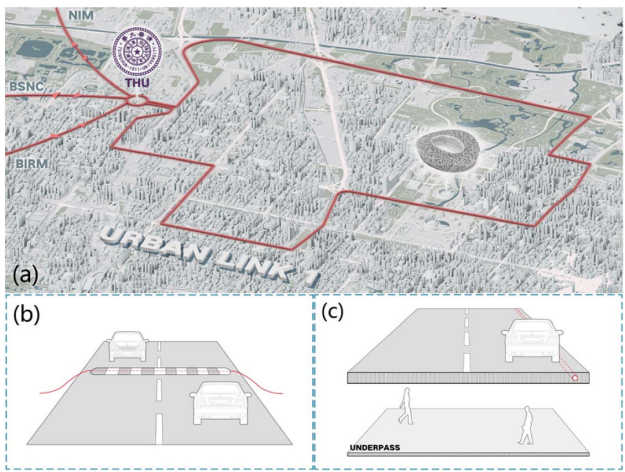

**Fig. 1.** Layout of the urban fiber test platform and scenarios of detected vibration events. (a) The fiber-based time-frequency network in Beijing mainly connecting Tsinghua University (THU) with the National Institute of Metrology (NIM), Beijing Satellite Navigation Center (BSNC), and Beijing Institute of Radio Metrology and Measurement (BIRM), respectively. To make a preliminary study of traffic monitoring, Urban Link 1 is chosen, which surrounds a relatively quiet area (the Olympic Stadium, education zones, and green parks). (b) The test scenario in campus. A part of armored fiber cable is placed through a speed bump of the road, from which we can localize vibrations of passing vehicles. (c) The test scenario on Urban Link 1. At the fourth ring road, a pedestrian underpass is in the way of the fiber link. Passing vehicles directly above this section generate vibrations which will be detected by the interferometry system.

window of the detected signal. The detailed discussion is provided in Appendix A.

In this paper, we propose the time shifting deviation (TSDEV) method to replace the commonly used cross-correlation method. Actually, the TSDEV method is more precise and has fewer boundary conditions. In principle, it can realize ultrahigh precision time delay estimation with a random cutoff window. In addition to the theoretical analysis, we set up a laser interferometer to carry out a lab test. The localization accuracy of ∼2.5 m is achieved and is mainly limited by the sampling rate (40 MS/s in our case). This result is more precise than that of the cross-correlation method. Furthermore, field tests in campus and on the urban link [Urban Link 1 in Fig. 1(a)] are carried out. Several interesting events are detected [Figs. 1(b) and 1(c)] and precisely localized. In campus, we localize the vibrations on a 9 m wide road and distinguish the cars running in opposite directions with 5.1 m localization difference. On the relatively quiet Urban Link 1, which surrounds the National Stadium (Bird's Nest) and the National Aquatics Center (Water Cube), we precisely localize interesting traffic vibrations whose changing rule is consistent with Beijing traffic regulation. We also propose a possible method which involves a more complicated network into the sensing ring to realize traffic monitoring in the future.

## 2. EXPERIMENTAL SETUP

The experimental setup of the laser interferometer is shown in Fig. 2(a). The interferometer consists of three parts: the optical system, sensing ring, and data processing part. The optical system transmits input laser signals into the sensing ring, receives output signals, and converts them into RF beat signals by the photodiode (PD). The sensing ring is the fiber link under detection; in our case, we set up an in-lab link, campus link, and urban link to analyze vibrations along them. In the data processing part, RF signals are acquired by the data acquisition (DAQ) system and transformed into phase information, from which we can extract vibration signals and localize them.

Starting from the laser source, the laser module we use is NKT Koheras BASIK X15, whose linewidth is <100 Hz and makes long distance interferometric detection viable. After the module, the laser is split into the reference beam and the signal beam. The signal beam passes through an acousto-optical modulator (AOM) and gets a frequency shifter $\nu_{AOM} = 107.8$ MHz to realize heterodyne configuration. Then, it is divided into two opposite directions. One travels in the sensing ring clockwise (CW beam) and the other counter-clockwise (CCW beam), with the help of two circulators. The outputs of the CW beam and the CCW beam will interfere with reference beams at PD 2 and PD 1, respectively. The DAQ system then converts them into digital signals. To avoid the phase error induced by the analog IQ demodulator [35], we utilize the digital IQ demodulation method to extract the phase information of signals. When vibration occurs in the sensing ring, the corresponding phase change will be induced on the CW beam and the CCW beam. It enables us to monitor the link's disturbance. In addition, two counter-propagating beams with phase change information will reach the corresponding PD with time delay $\tau_0$, and make it possible to localize the vibration events [36–38].

## 3. METHODS

### A. Vibration Localization Principle

The vibration localization principle is illustrated in Fig. 2(b). The optical path length between the vibration point and PD 1 is $\Delta_1$. Thus, the CCW beam with phase change information will reach PD 1 at $t_1 = t_0 + \Delta_1/c$, where $t_0$ is the time when vibration happens. The optical path length between the vibration point and PD 2 is $\Delta_2$, which means the CW beam with phase change information will reach PD 2 at $t_2 = t_0 + \Delta_2/c$. The overall optical path of the sensing ring $\Delta = \Delta_1 + \Delta_2$ can be calibrated in advance [39,40].

Using the time delay estimation method, we can extract the phase change time delay $\tau_0 = (\Delta_2 - \Delta_1)/c$ of the CW and CCW beams. Then, we can calculate the optical path length $\Delta_1$ and localize the vibration point:

$$\Delta_1 = \frac{1}{2}(\Delta - c \cdot \tau_0).  \qquad (1)$$

Equation (1) shows that localization accuracy mainly depends on the accuracy of $\tau_0$. Two key factors will influence the estimation of $\tau_0$. One is the sampling rate of the DAQ system. In our case, the 40 MS/s sampling rate leads to 3.75 m optical path length resolution, which equals 2.5 m fiber length resolution approximately (considering the fiber refractive index of 1.5). However, to achieve the resolution limitation above, the time delay estimation method as the other factor should be precise enough. The commonly used cross-correlation method has nonnegligible random error when dealing with practical signals (time-finite signal segments) and can only approach the resolution limitation in some specific conditions (more details available in Appendix A). We directly draw our conclusion of the cross-correlation method: only if two signal segments have the same picture as each other [41], can we obtain the precise estimation result without theoretical error. This condition rarely attracts attention, because in previous works, the localization accuracy of the forward-transmission scheme can hardly reach meter magnitude. But in the case of ultrahigh precision localization, the residual error of the cross-correlation method will appear as the major source of localization error.

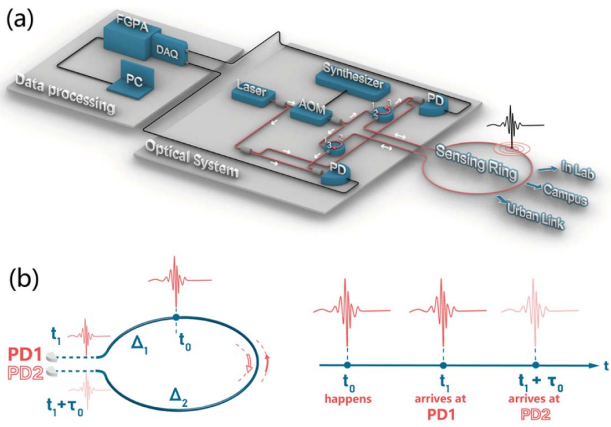

**Fig. 2.** Experimental setup of laser interferometer and the schematic diagram of vibration localization principle. (a) The vibration detection system consists of three parts: optical system, sensing ring, and data processing part. Optical system: laser (1550 nm laser module), acousto-optical modulator (AOM), synthesizer (providing shifter frequency for AOM), photodiode (PD). Sensing ring: in-lab test, campus test, and urban link test are carried out, respectively. Data processing part: data acquisition (DAQ) system, field programmable gate array (FPGA), computer for data processing (PC). (b) The principle of vibration localization is using the counter-propagating beams to detect the vibration event happening at $t_0$. The optical path lengths between the vibration point and PD 1, PD 2 are $\Delta_1$ and $\Delta_2$, respectively. Thus, the arrival times of two beams with vibration caused phase change are $t_1 = t_0 + \Delta_1/c$ and $t_2 = t_0 + \Delta_2/c$. The time delay $\tau_0 = t_2 - t_1$ can be used to localize the vibration.

### B. Time Shifting Deviation Method

Actually, no matter how we improve the cross-correlation method, the estimation error will always exist [42]. Thus, we propose a novel time delay estimation method, TSDEV. Instead of the correlation coefficient in the cross-correlation method, we use the standard deviation of two signals' difference $\sigma[x_1(t) - x_2(t + \tau)]$ as the estimation indicator:

$$\mathrm{TSDEV}(\tau)$$
$$= \sigma[x_1(t) - x_2(t + \tau)]$$
$$= \sqrt{\frac{1}{T}\int_{t_w}^{t_w+T}\{x_1(t) - x_2(t+\tau) - E[x_1(t) - x_2(t+\tau)]\}^2 \mathrm{d}t}.$$
(2)

In Eq. (2), $x_1(t)$ and $x_2(t)$ are the detected phase changes of the CCW beam and the CW beam, respectively. $\sigma[x_1(t) - x_2(t + \tau)]$ is the standard deviation of their difference, and $E[x_1(t) - x_2(t + \tau)]$ is its expectation. During the integration calculation of the time-finite signal segments, $t_w$ is the initial time of the cutoff window, and $T$ is the window size. It is obvious that only if $x_1(t)$ strictly equals $x_2(t + \tau)$, the $\mathrm{TSDEV}(\tau) = 0$ will happen. In any other situations, the $\mathrm{TSDEV}(\tau) \geq 0$ will always occur, which means $\mathrm{TSDEV}(\tau)$ reaches the minimum when $x_1(t) = x_2(t + \tau)$.

In the localization case, $x_1(t)$ and $x_2(t)$ are both caused by the same vibration, but have time delay $\tau_0$ due to different optical path lengths; thus, an approximation $x_2(t) = x_1(t - \tau_0)$ can be made when noise is ignored. Therefore, the $\mathrm{TSDEV}(\tau)$ will be minimum only if time shifting $\tau$ meets the real time delay $\tau_0$. When we calculate $\mathrm{TSDEV}(\tau)$ and find out its minimum point, we will get the precise result $\tau_0$ and the corresponding position $\Delta_1$.

TSDEV is a more rigorous method than cross correlation. To make a further analysis of TSDEV, we carry out a calculation in Eq. (3). The phase changes $x_1(t)$ and $x_2(t)$ are set as sine wave with frequency $\omega$, and time delay $\tau_0$ to be zero. Then, $\mathrm{TSDEV}(\tau)$ is as follows:

$$\begin{aligned}\mathrm{TSDEV}^2(\tau) &= \frac{1}{T}\int_{t_w}^{t_w+T}[x_1(t) - x_2(t+\tau) - C(\tau)]^2 \mathrm{d}t \\ &= \frac{1}{T}\int_{t_w}^{t_w+T}\{\sin(\omega t) - \sin[\omega(t+\tau)] - C(\tau)\}^2 \mathrm{d}t \\ &= D(\tau)\cdot\sin^2\left(\frac{1}{2}\omega\tau\right) + G(\tau)\cdot\sin\left(\frac{1}{2}\omega\tau\right) + C^2(\tau),\end{aligned}$$
(3)

where $C(\tau) = \frac{1}{T}\int_{t_w}^{t_w+T}[x_1(t) - x_2(t+\tau)]\mathrm{d}t$ is the mean value of the two signals' difference, $D(\tau) = \frac{1}{\omega T}[2\omega T + \sin(2\omega t_w + 2\omega T + \omega\tau) - \sin(2\omega t_w + \omega\tau)]$, and $G(\tau) = \frac{4C(\tau)}{\omega T}[\sin(\omega t_w + \omega T + \frac{1}{2}\omega\tau) - \sin(\omega t_w + \frac{1}{2}\omega\tau)]$, respectively.

From Eq. (3), we can know that only if $\tau = 0$, will $\mathrm{TSDEV}(\tau) = 0$ occur. When the value of $\tau$ changes, $\mathrm{TSDEV}(\tau) > 0$ is always the case due to the property of standard deviation. Unlike the result of cross correlation (full details are provided in Appendix A), the accuracy of TSDEV estimation is not related to window size $T$ of signal segments. That is to say, the estimation result is random error free for any cutoff of phase change signals.

In addition, the influence of background noise $n(t)$ is also considered, and we find that a modified TSDEV method can still work out the localization results precisely. The method is using the TSDEV of background noise ($\sigma[n_1(t) - n_2(t + \tau)]$) to compensate the TSDEV of detected signals, which can decrease the influence of noise and reach the limitation of the sampling rate. Full details are provided in Appendix B.

## 4. RESULTS

### A. In-Lab Demonstration

To demonstrate the advantages of the TSDEV method, we carry out an in-lab experiment and make a comparison between the results of TSDEV and the cross-correlation method. Fiber spools are utilized to form the sensing ring, as shown in Fig. 3(a). One is around 50 km (49.49 km), and the other spool $L$ is set to be around 10 km and 50 km, respectively, to explore the localization accuracy when the sensing ring has different lengths. In the middle of the two spools, a fiber stretcher (FST) is employed to provide controllable vibration.

First, a 50–10 km scheme is under test ($L$ is around 9.84 km). The corresponding phase change of the CW and CCW beams is obtained and shown in Fig. 3(b), from which the time delay $\tau_0$ can be extracted. To make it a more intuitive representation, the localization result is transformed from the optical path length to the fiber length, where the refractive index of the fiber is roughly set as 1.5. Thus, we localize vibration at the point 49,493.5 m away from the reference point using the TSDEV method. The window size of signal segments is randomly chosen and will not influence the result. After repeating the trials, the standard deviation of the measurement results is 2.4 m and all data fall into the error range of ±5 m, as shown in Fig. 3(c). Actually, the localization accuracy nearly reaches the limitation of sampling rate 40 MS/s. It can also be demonstrated by the fitted probability density curve, in which the fitting width of 1 standard deviation (1 Std corresponding to probability 68.3 %) is 2.5 m. The detailed probability distribution curves are shown in Fig. 3(d).

Considering the cross-correlation method, we use different window sizes to localize the vibration mentioned above. Since the FST-induced vibration is single frequency, we cut off signal segments with integer multiples of half-period (Cutoff 1: 26 half-periods) and random length (Cutoff 2: 26.04 half-periods), respectively. Their results are shown in Fig. 3(c) to compare with the results of the TSDEV method. For those segments with Cutoff 1, the average position is 49,493.0 m away from the reference point and the standard deviation is 3.6 m. All results are centered within an error range of ±12.5 m, slightly larger than that of the TSDEV method. However, when signals are cut off with random length (Cutoff 2), the localization error increases soon. The average position slides to 49,482.3 m away from the reference point with a standard deviation of 5.7 m, and the error range grows to ±17.5 m. It demonstrates that the cross-correlation method has nonnegligible random error and can only approach the accuracy of the TSDEV method in some specific conditions. When target vibration signals are pure sine waves, this condition is cutting off signals with integer multiples of half-period. If taking arbitrary signals into account, the condition will be more tightened, which is analyzed in Appendix A.

Then, we extend the sensing ring to the 50–50 km scheme ($L$ is around 48.22 km). The result using the TSDEV method shows that vibration occurs at 49,493.3 m away from the reference point, and the standard deviation is 2.2 m. Full details are provided in Appendix C. This average point and standard deviation are both consistent with that of the 50–10 km scheme, which means the 50–50 km scheme also reaches the resolution limitation, and the laser interferometer has good reproducibility. In Fig. 3(d), we display the probability distribution curve of two schemes. The confidence intervals' widths corresponding to the 1, 2, 3 Stds are listed in Fig. 3(d). In our system, the length of the sensing ring is within 100 km, and the optical amplifier is not used. The Rayleigh scattering is much weaker comparing with signal light, and the coherent Rayleigh interference is not the main noise source in the localization process. For a longer sensing ring, the Rayleigh scattering can be efficiently suppressed via adding two AOMs with different frequency shifting on two counter-propagating signal beams [43,44].

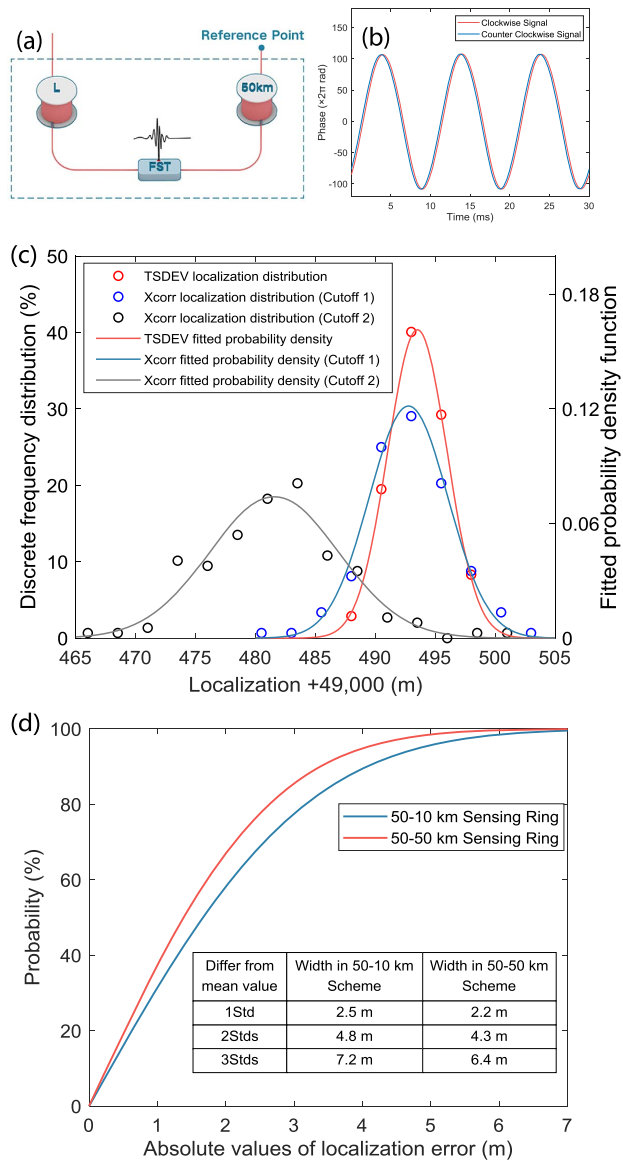

**Fig. 3.** Diagram of the sensing ring and detection results of in-lab demonstration. (a) The sensing ring consists of a 50 km fiber spool, a fiber stretcher (FST), and a fiber spool $L$. (b) The phase change detected by the CCW beam (blue) and the CW beam (red). (c) The localization results of the 50–10 km scheme, in which the hollow circle points represent the discrete distribution frequency and the lines are the fitted probability density curve. Red plots: using the TSDEV method. Blue plots: using the cross-correlation (Xcorr) method with signal segments of Cutoff 1. Black plots: using the Xcorr method with signal segments of Cutoff 2. The measurement in the 50–10 km test is repeated 277 times. (d) The fitted probability distribution curves of the 50–10 km (blue) and the 50–50 km (red) scheme results (using the TSDEV method). The table inside shows the widths of the fitting curve corresponding to the 1, 2, and 3 Stds. The measurement in the 50–50 km test is repeated 175 times.

### B. Field Test in Campus

To further demonstrate the superiority of the TSDEV method, we carry out a field test in the campus of Tsinghua University. As shown in Fig. 4(a), we connect the sensing ring to a part of the fiber cable in Tsinghua Campus, analyze vibration events induced by cars passing through the cable and localize them. The sensing ring consists of three parts: 50 km (49.49 km) fiber spool in lab, 800 m fiber cable in campus, and 10 km (9.84 km) fiber spool in lab. A piece of the 800 m armored fiber cable is placed through a speed bump.

When cars drive across it, corresponding phase change [as shown in Fig. 4(b)] will allow us to extract and localize the vibrations. It is worth mentioning that the width of the road is 9 m, and cars from opposite directions (driving on the right) lead to different localization results. We divide the collected data into two parts in terms of driving direction and show localization results in Fig. 4(c). The average position of Direction 1 is 49,825.3 m away from the reference point, and the standard deviation is 8.4 m. The average position of Direction 2 is 49,830.4 m away from the reference point. Its standard deviation is 6.6 m. Obviously, the localized position of Direction 1 is closer to the reference point than that of Direction 2, which is in line with the setup of the sensing ring configuration in Fig. 4(a). The bias value 5.1 m between two center positions is also reasonable, considering the 9 m road width. While using the cross-correlation method, we cannot distinguish the vibrations in these two directions, and full details are provided in Appendix D.

### C. Field Test on the Urban Link

To study the traffic cases in a larger district, we choose an urban fiber link [Fig. 5(a)] surrounding a relatively quiet area as the sensing ring. The link is buried underground deeply. It has a length of 31.42 km and passes through Olympic center, green parks, and education zones. As a preliminary study, this specific selection makes the results easy to analyze. We also connect a 10 km (9.84 km) fiber spool with Urban Link 1 as extension and make the sensing ring's total length be 41.26 km.

In the top part of Fig. 5(b), the phase information shows us an interesting vibration sequence. Before midnight (21:00–23:00), vibrations behave as low-amplitude ($<200\pi$ rad phase change peak-to-peak) and high-frequency (10–20 Hz) fluctuations. After midnight (24:00–3:00), vibrations change to be shockwaves with high amplitude ($400\pi$–$1200\pi$ rad peak-to-peak) and relatively low frequency (~4 Hz). The typical phase changes in these two periods are shown in the middle part of Fig. 5(b). All these vibrations perform good consistency in localization results using the TSDEV method. We extract 50 vibration segments and localize them. Their average position is at 15,749.0 m away from Tsinghua University (clockwise along Urban Link 1). An 18.5 m standard deviation demonstrates that these vibrations occur at the same position, and full details are provided in Appendix E.

With the help of the urban fiber route diagram, we find out the location of the vibration source and discover a pedestrian underpass there, as shown in Fig. 5(a). At that point, urban fiber cable is deployed above the underpass and beneath the fourth ring road, much more shallowly than other places. Thus, the vibrations caused by passing vehicles can transmit to fiber cable without much attenuation. In addition, the soil between the ring road and the underpass forms a resonant cavity, which increases the detection sensitivity to some extent.

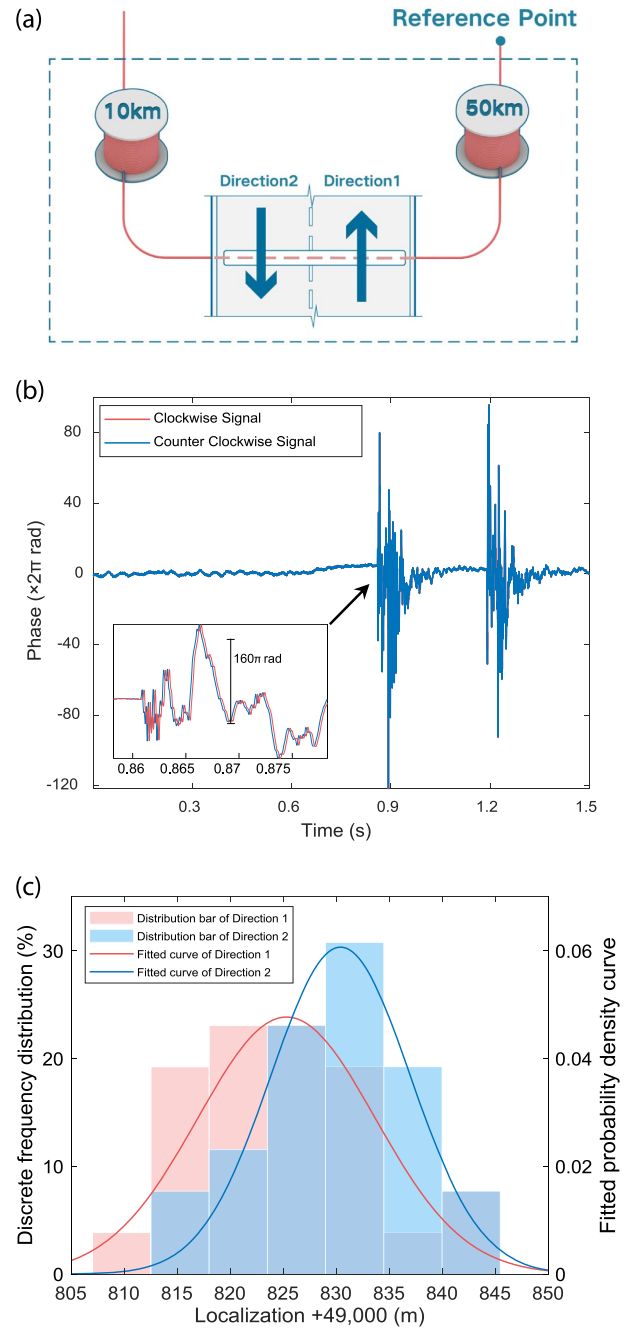

**Fig. 4.** Diagram of the sensing ring and detection results of the campus test. (a) The sensing ring consists of a 50 km fiber spool in lab, an 800 m fiber cable in campus, and a 10 km fiber spool in lab. A 9 m part of the 800 m fiber cable is placed through the speed bump of the campus road to detect passing vehicles running in Direction 1 and Direction 2. (b) The phase change detected by the CCW beam (blue) and the CW beam (red), with time delay $\tau_0$. The two envelopes correspond to vibrations caused by front and rear wheels of passing vehicles. (c) The localization results of the campus test. Red bars: the distribution frequency of Direction 1. Red line: the fitted probability density curve. Blue bars: the distribution frequency of Direction 2. Blue line: the fitted probability density curve. The measurement in the campus test is repeated 56 times.

By monitoring the traffic at that point, the changing rule of vibrations is revealed. As shown in the bottom part of Fig. 5(b), we display two sketch maps which simply present the traffic flow cases. Before midnight, vehicles on the road are mainly cars and jeeps, which are light-weight and fast-running (corresponding to low-amplitude and high-frequency vibrations).

After 23:00, big lorries are permitted to enter the downtown city; the corresponding regulation is provided in Appendix E. The lorries generate strong vibrations which lead to the length change of fiber cable up to 600 μm (1200π rad phase change peak-to-peak typically). Since they have a longer body, their vibrations have relatively low frequency (4 Hz typically). After 24:00, the lorries' flow rate climbs to the peak. Their corresponding vibrations are continuous and appear as a high-amplitude base band in the top part of Fig. 5(b). In conclusion, the changing rule of traffic flow is consistent with the amplitude and frequency of detected vibrations, which is the evidence that the laser interferometer with the TSDEV method is highly precise and qualified in urban traffic localization cases. Via the time-frequency fiber network, the proposed system has the potential to form a large-scale sensing network and localize traffic events with ultrahigh precision.

## 5. DISCUSSION

We obtain vibration information from the laser interferometer and use the TSDEV method to localize vibration events. The TSDEV method is more precise than the cross-correlation method. It has fewer boundary conditions, which leads to advantages in localization accuracy and practicability. The in-lab experiment demonstrates that localization accuracy is limited by the sampling rate of 40 MS/s. The standard deviations of localization results are ∼2.5 m on both the 50–10 km and 50–50 km schemes. Field tests are also carried out in campus and on the urban fiber link. We localize the passing vehicles on a 9 m road in Tsinghua Campus and recognize the 5.1 m location difference between cars from opposite directions. On Urban Link 1, we localize a vibration source on the fourth ring road. By monitoring the traffic flow there, we find out that the changing rule of vibrations is consistent with traffic conditions. The standard deviation of localizing results is different in the in-lab test, campus test, and urban link test. It worsens from ∼2.5 m to 8.4 m, and finally to 18.4 m. This is mainly because the vibration range of the fiber link gets larger and larger in these situations. In the lab test, we use the FST to generate vibration. The FST is placed at a fixed position of the fiber link. While in the campus test, an ∼9 m fiber cable will be affected by passing vehicles. On the urban field test, vibrations happen around an underpass of the fourth ring road, which is much wider than the campus road. Thus, larger and larger standard deviations are reasonable and do not mean a loss in localization accuracy.

Initially, we want to choose a quiet urban fiber link to detect the background noise. Apart from the underpass, Urban Link 1 hardly feels vibrations on the rest of the link. Thus, vibrations at the underpass dominate and nearly submerge others. If we want to carry out further analysis, data processing such as filtering and signal extraction should be made in advance.

Further investigation will be focused on how to localize vibrations which occur at different positions of the link simultaneously. A procedure may be used to solve this problem. When several vibrations occur simultaneously, the minimum value of TSDEV will be larger than that of a single vibration. This could be an indicator to judge the number of vibrations. Then, the signal can be separated into different frequency bands. Because detected vibrations are actually broadband signals, we can always find a frequency band in which a single vibration dominates. TSDEV will be carried out in each frequency band, and the minimum values of TSDEV can help to judge whether it is a single vibration. If so, the corresponding time delay can be used to localize this vibration source. After going through all frequency bands, the localization of multiple vibrations can be realized.

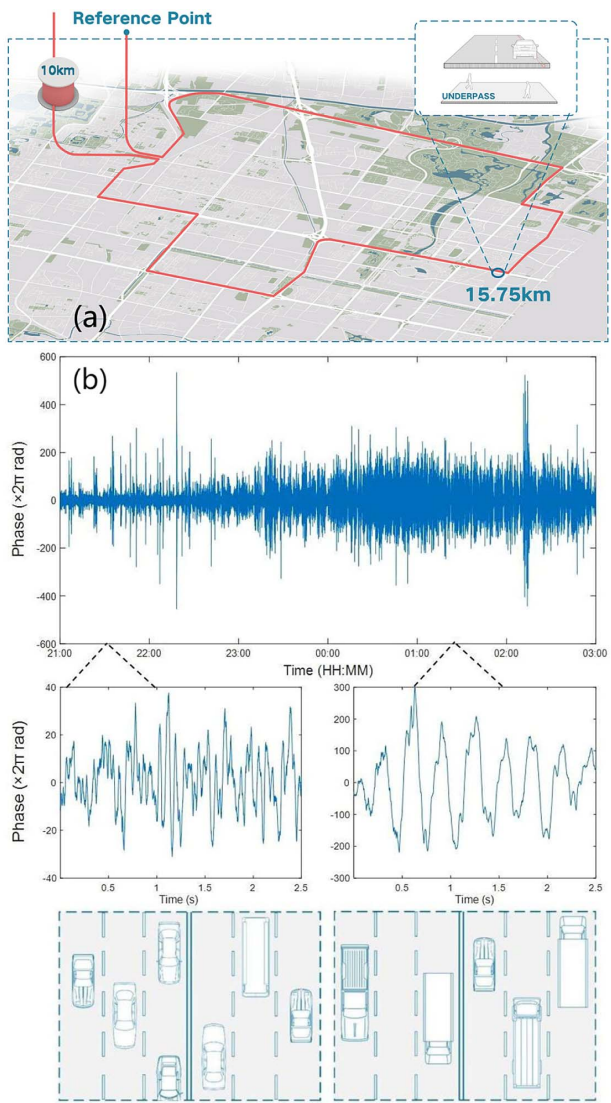

**Fig. 5.** Diagram of the sensing ring and detection results of the urban link test. (a) The sensing ring consists of a 10 km fiber spool in lab and a 31.42 km Urban Link 1. A vibration source (pedestrian underpass) is localized at point 15.749 km away from Tsinghua Campus, clockwise along the link. (b) Top part: the detected phase change from 21:00 to 3:00. Middle part: two segments of typical vibrations before 23:00 and after 24:00. Bottom part: sketch maps of traffic flow before and after 23:00.

## APPENDIX A: ERROR ANALYSIS OF THE CROSS-CORRELATION METHOD

The commonly used cross-correlation method utilizes the expectation of $x_1(t) \times x_2(t + \tau)$ as the indicator to estimate the time delay $\tau_0$ between two signals. However, the detected phase changes [$x_1(t)$ and $x_2(t)$] have finite length in practice and cannot meet the conditions of theoretical calculation [41]. Thus, localization error induced by the cross-correlation method is inevitable and will bring nonnegligible bias in ultrahigh precision localization cases.

A simplified calculation is carried out as a preliminary demonstration, when we set the phase change to be pure sine wave:

$$\begin{aligned} x_{\text{corr}}(\tau) &= \frac{1}{T} \int_{t_w}^{t_w+T} x_1(t) \cdot x_2(t+\tau) dt \\ &= \frac{1}{T} \int_{t_w}^{t_w+T} \sin(\omega t) \cdot \sin[\omega(t+\tau)] dt \\ &= \frac{1}{2} \cos(\omega\tau) - \frac{1}{2\omega T} \sin(\omega T) \cdot \cos(2\omega t_w + \omega T + \omega\tau) \end{aligned}$$ (A1)

In Eq. (A1), $x_1(t)$ and $x_2(t)$ are the detected phase changes with time delay $\tau_0$, $t_w$ is the initial moment of two signals' cutoff window, and $T$ is the window size of the cross-correlation segments. Specifically, we can set the frequency of phase changes to be $\omega$ and the time delay $\tau_0$ to be zero. Thus, the precise estimation result should be $\tau = 0$, which means $x_{\text{corr}}(\tau)$ should reach maximum when $\tau = 0$.

However, the derivative of $x_{\text{corr}}(\tau)$ is also influenced by window size $T$. To calculate the maximum position, we set $x_{\text{corr}}'(\tau) = 0$ and obtain the equation of $\tau$:

$$\tan(\omega\tau) = \frac{\sin(\omega T) \cdot \sin(2\omega t_w + \omega T)}{\omega T - \sin(\omega T) \cdot \cos(2\omega t_w + \omega T)}.$$ (A2)

In Eq. (A2), to meet the ideal situation $\tau = 0$, the right-hand side must equal zero. Since $t_w$ can be randomly chosen, the term $\sin(\omega T)$ should be zero, which means window size $T$ should meet the condition $T = N \cdot (\pi/\omega)$, $N = 1, 2, 3, \ldots$. Thus, when phase changes are sine wave, only if we cut off the signal with integer multiples of half-period, can we obtain precise estimation results using the cross-correlation method.

The conclusion above will be more tightened when considering the normal case, in which phase changes are arbitrary signal segments. For time-finite signals $s_1(t) = x_1(t)|_{t\in(t_w, t_w+T)}$ and $s_2(t) = x_1(t - \tau_0)|_{t\in(t_w, t_w+T)}$, the two signals are both included in segment $s(t) = x_1(t)|_{t\in(t_w-\tau_0, t_w+T)}$. Thus, we can treat $s(t)$ as one period of a periodic signal (similar to the operation of the discrete Fourier transform [45]) and transform it into Fourier series expansion (FS). At that time, the fundamental period of the FS should be $T + \tau_0$, and its sine series can be written as

$$s(t) = a_0 + \sum_{n=1}^{\infty} a_n \sin(n\omega_0 t + \varphi_n), \quad t \in (t_w - \tau_0, t_w + T).$$ (A3)

The fundamental frequency $\omega_0$ is determined by the window size $T + \tau_0$, with relation $\omega_0 = 2\pi/(T + \tau_0)$. Since $s_1(t)$ and $s_2(t)$ are different cutoffs of $s(t)$, they can be written as the same sine series:

$$s_1(t) = a_0 + \sum_{n=1}^{\infty} a_n \sin(n\omega_0 t + \varphi_n), \quad t \in (t_w, t_w + T),$$ (A4)

$$s_2(t) = a_0 + \sum_{n=1}^{\infty} a_n \sin[n\omega_0(t - \tau_0) + \varphi_n],$$
$$t \in (t_w, t_w + T).$$ (A5)

It should be noted that the effective length of the sine series is $(t_w, t_w + T)$, corresponding to the initial window lengths of $s_1(t)$ and $s_2(t)$. When the cross-correlation method is used, it should be integrated within $(t_b, t_b + T_1)$, where $t_w \leq t_b$, $T_b \leq T$:

$$\begin{aligned} x_{\text{corr}}(\tau) &= \frac{1}{T_b} \int_{t_b}^{t_b+T_b} s_1(t) \cdot s_2(t+\tau) dt \\ &= \frac{1}{T_b} \int_{t_b}^{t_b+T_b} \left[ a_0 + \sum_{n=1}^{\infty} a_n \sin(n\omega_0 t + \varphi_n) \right] \\ &\quad \cdot \left\{ a_0 + \sum_{n=1}^{\infty} a_n \sin[n\omega_0(t - \tau_0 + \tau) + \varphi_n] \right\} dt \\ &= \frac{1}{T_b} \int_{t_b}^{t_b+T_b} \sum_{k=0}^{\infty} \sum_{n=0}^{\infty} \{a_k \sin(k\omega_0 t + \varphi_k) \\ &\quad \cdot a_n \sin[n\omega_0(t - \tau_0 + \tau) + \varphi_n]\} dt. \end{aligned}$$ (A6)

In Eq. (A6), we can divide the integral terms into two cases, when $k = n$ and $k \neq n$. The situations when $k = n$ are already analyzed in Eqs. (A1) and (A2). When $k \neq n$, the corresponding $x_{\text{corr}}(\tau)$ is as follows:

$$\begin{aligned} x_{\text{corr}}(\tau) &= \frac{1}{T_b} \int_{t_b}^{t_b+T_b} a_k \sin(k\omega_0 t + \varphi_k) \\ &\quad \cdot a_n \sin[n\omega_0(t - \tau_0 + \tau) + \varphi_n] dt \\ &= \frac{a_k a_n}{T_b} \{A(T_b) \cos[n\omega_0(\tau_0 - \tau)] \\ &\quad + B(T_b) \sin[n\omega_0(\tau_0 - \tau)]\}, \end{aligned}$$ (A7)

where $A(T_b) = \frac{\sin[(k-n)\omega_0 T_b/2] \cdot \cos[(k-n)\omega_0 t_b + (k-n)\omega_0 T_b/2 + \varphi_k - \varphi_n]}{(k-n)\omega_0}$
$- \frac{\sin[(k+n)\omega_0 T_b/2] \cdot \cos[(k+n)\omega_0 t_b + (k+n)\omega_0 T_b/2 + \varphi_k + \varphi_n]}{(k+n)\omega_0}$, and
$B(T_b) = -\frac{\sin[(k-n)\omega_0 T_b/2] \cdot \sin[(k-n)\omega_0 t_b + (k-n)\omega_0 T_b/2 + \varphi_k - \varphi_n]}{(k-n)\omega_0}$
$- \frac{\sin[(k+n)\omega_0 T_b/2] \cdot \sin[(k+n)\omega_0 t_b + (k+n)\omega_0 T_b/2 + \varphi_k + \varphi_n]}{(k+n)\omega_0}$, respectively.

Ideally, $x_{\text{corr}}(\tau)$ in Eq. (A6) should reach maximum when $\tau = \tau_0$. We set $x_{\text{corr}}'(\tau) = 0$ and obtain the corresponding $\tau$:

$$\tan[n\omega_0(\tau_0 - \tau)] = \frac{B(T_b)}{A(T_b)}, \quad n = 0, 1, 2.\ldots \quad \text{(A8)}$$

From Eqs. (A7) and (A8), we can find out that maximum position of $x_{\text{corr}}(\tau)$ is determined by $T_b$, the window size of cross-correlation integration. Only if $T_b = T + \tau_0$ occurs, the $x_{\text{corr}}(\tau)$ for different frequency ($k \neq n$) is zero, and $x_{\text{corr}}(\tau)$ for the same frequency ($k = n$) reaches the maximum at $\tau = \tau_0$. However, as mentioned before, the effective length of the sine series is $(t_w, t_w + T)$. Therefore, there will always be estimation error in the cross-correlation method unless a specific condition is met. We find out that when all the information of two segments is included in $s_1(t)$ [$x_1(t)|_{(t_w+T-\tau_0, t_w+T)} = x_1(t)|_{(t_w-\tau_0, t_w)}$], the signal extension is not necessary [$s(t) = s_1(t)$] and the period of the Fourier series is $T$. At that time, $x_{\text{corr}}(\tau)$ can reach maximum at $\tau = \tau_0$ when the cutoff length $T_b = T$, which means a precise estimation is obtained via the cross-correlation method.

## APPENDIX B: MODIFIED TSDEV METHOD TO DECREASE THE INFLUENCE OF BACKGROUND NOISE

In Section 3, we demonstrate the ultrahigh precision of the TSDEV method by theoretical analysis. Besides the ideal case $x_2(t) = x_1(t - \tau_0)$, where $x_1(t)$ and $x_2(t)$ are the detected phase change signals, the influence of background noise is also taken into account. A modified TSDEV method is proposed to compensate the error induced by noise and can work out the localization results precisely.

To make it clear, we separate background noise into two parts: related noise $n_s(t)$ and random noise $n_r(t)$. Since two signals are propagating in the same detection system, they have related noise $n_s(t)$. On the other hand, due to the random noise of instruments, two signals will have different noise $n_{r1}(t)$ and $n_{r2}(t)$. Thus, the signals can be written as

$$s_1(t) = x_1(t) + n_s(t) + n_{r1}(t),$$
$$s_2(t) = x_1(t - \tau_0) + n_s(t - \tau_s) + n_{r2}(t). \quad \text{(B1)}$$

In Eq. (B1), $\tau_s$ is the time delay between the two signals' related noise, and it is caused by the system's asymmetry. Since $\tau_s \neq \tau_0$, the accuracy of TSDEV will be influenced by both $n_s(t)$ and $n_r(t)$.

Hence, we propose a compensation method. We choose a segment of background noise near the signals, which can be written as $s'(t') = n_s(t') + n_r(t')$, and compensate the TSDEV of $s(t)$ by that of $s'(t')$. This modified TSDEV method has two bases. First, the statistic law of $n_s(t)$ changes slowly compared with the duration of vibration detection, and thus $n_s(t')$ can be used to compensate $n_s(t)$. Second, vibration and background noise are uncorrelated, which makes it possible to separate their standard deviations as follows:

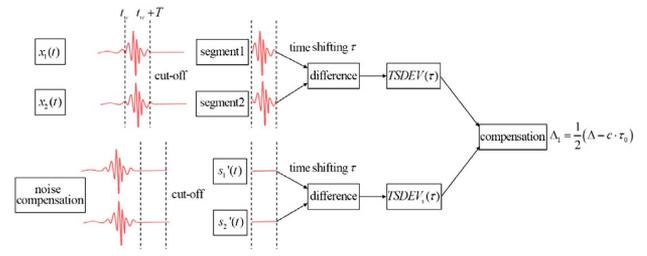

**Fig. 6.** Diagram of the localizing procedure, using the demodulated phase change signals $x_1(t)$ and $x_2(t)$ to localize the vibration $\Delta_1$.

$$\text{TSDEV}(\tau)$$
$$= \sigma[s_1(t) - s_2(t + \tau)] - \sigma[s'_1(t') - s'_2(t' + \tau)]$$
$$= \{\sigma[x_1(t) - x_1(t - \tau_0 + \tau)] + \sigma[n_s(t) - n_s(t - \tau_s + \tau)]$$
$$+ \sigma[n_{r1}(t)] + \sigma[n_{r2}(t + \tau)]\}$$
$$- \{\sigma[n_s(t') - n_s(t' - \tau_s + \tau)] + \sigma[n_{r1}(t')] + \sigma[n_{r2}(t' + \tau)]\}$$
$$\approx \sigma[x_1(t) - x_1(t - \tau_0 + \tau)]. \quad \text{(B2)}$$

When background noise cannot be ignored, the TSDEV of background noise can be used to compensate the noise-induced error, and a precise estimation of time delay $\tau_0$ can be obtained. This theory can support us to localize vibrations with raw data even if the noise level is relatively high. It also decreases the danger of information loss during data processing such as filtering and transforming.

In conclusion, the vibration localizing procedure is as follows. First, we demodulate the two counter-propagating beams and obtain phase change signals $x_1(t)$ and $x_2(t)$. When vibration occurs, vibration signal segments can be extracted from $t_w$ to $t_w + T$, where the duration of vibration is larger than $T$. Then, we can calculate the difference between the two signals' segments $[x_1(t) - x_2(t + \tau)]|_{t_w}^{t_w+T}$, obtain the TSDEV function TSDEV($\tau$) via continuous time shifting as shown in Eq. (2), and get the time delay $\tau_0$ between two vibration signals. Finally, Eq. (1) is used to localize the vibration $\Delta_1$. When the localizing results are influenced by background noise, we can extract signal segments from the background noise and calculate its TSDEV function TSDEV$_s(\tau)$ to compensate TSDEV($\tau$). The diagram of this procedure is shown in Fig. 6.

## APPENDIX C: LOCALIZATION RESULTS OF 50–50 km SCHEME IN-LAB DEMONSTRATION

In Section 4, we carry out an in-lab demonstration to prove that the localization accuracy reaches the resolution limited by the sampling rate.

The results of the 50–50 km scheme are shown in Fig. 7 as supplementary information for our demonstration. The refractive index of the fiber is roughly set as 1.5, so that the optical path length can be transformed into the fiber length, which is more convenient to be understood. We localize the vibration at 49,493.3 m position away from the reference point, and the standard deviation is 2.2 m. All data fall into the error range of ±5 m, as shown in Fig. 7. From the fitted probability density curve, the width of 1 standard deviation (1 Std, correspond-ing to probability 68.3%) is ±2.2 m, which proves that the localization accuracy of the 50–50 km sensing ring has reached the 2.5 m resolution limitation.

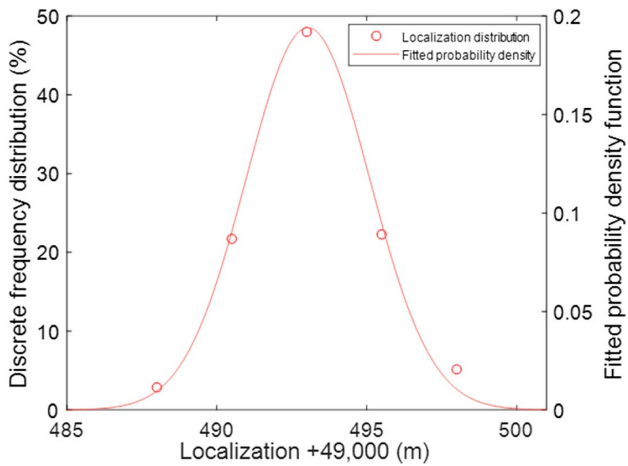

**Fig. 7.** Detection results of the 50–50 km scheme in-lab demonstration. The red hollow circle points are the distribution frequency using the TSDEV method, and the red line is the fitted probability density curve.

## APPENDIX D: LOCALIZATION RESULTS OF THE FIELD TEST IN CAMPUS USING THE CROSS-CORRELATION METHOD

In Section 4, we deploy a piece of armored fiber cable through a speed bump in the campus of Tsinghua University. The passing vehicles can be detected and localized via the TSDEV method, and cars from opposite directions lead to a localization difference of 5.1 m. By contrast, we also use the cross-correlation method to deal with the same phase change information. The localization results are shown in Fig. 8. The average localizing position has a deviation of ∼29 m, and the standard deviation is 53.9 m and 46.8 m, for cars driving from Direction 1 and Direction 2, respectively. This accuracy is much worse than that of TSDEV (8.4 m and 6.6 m), which demonstrates the novelty of the TSDEV method again.

## APPENDIX E: LOCALIZATION RESULTS OF THE FIELD TEST ON URBAN LINK 1

In Section 4, we choose Urban Link 1 as a part of the sensing ring to carry out the field test on the urban fiber link. We discover a series of vibrations which have good consistency in localization results using the TSDEV method. We extract 50 segments of vibration signals and localize them. The histogram of their discrete frequency distribution is shown in Fig. 9, from which we can find out that the average point is 15,749.0 m away from Tsinghua Campus, along Urban Link 1 counter-clockwise. The standard deviation of results is 18.5 m, small enough to prove that all these vibrations occur at the same position on the link. By contrast, the localizing results using the cross-correlation method will deviate from the underpass location several kilometers.

In addition, changing rule of the detected vibrations is also associated with the traffic regulation. In terms of the notification published by the Beijing Traffic Management Bureau: "Notification on the traffic control of a part of lorries for the reduction of pollutant release," cargo lorries are forbidden to enter the downtown city of Beijing (inside the fifth ring road) during 6:00–23:00. Thus, at the localized point (near an exit of the north fourth ring road), light-weight cars are the main part of traffic before midnight, as shown in Fig. 10(a), and produce low-amplitude high-frequency fluctuations. After 23:00, big lorries are permitted to enter the downtown city, as shown in Fig. 10(b), which generates shockwaves with high amplitude and relatively low frequency. This phenomenon is in line with the traffic notification, which provides a reasonable interpretation of vibrations' amplitude and frequency change and proves the localization correctness from another aspect.

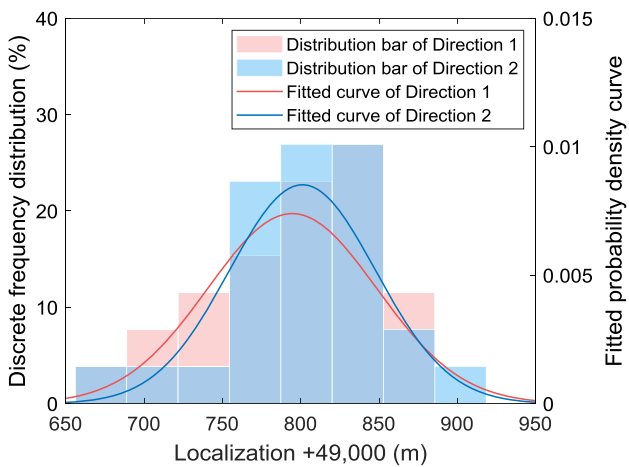

**Fig. 8.** Localization results of the campus test using the cross-correlation method. Red bars: the distribution frequency of Direction 1. Red line: the fitted probability density curve. Blue bars: the distribution frequency of Direction 2. Blue line: the fitted probability density curve.

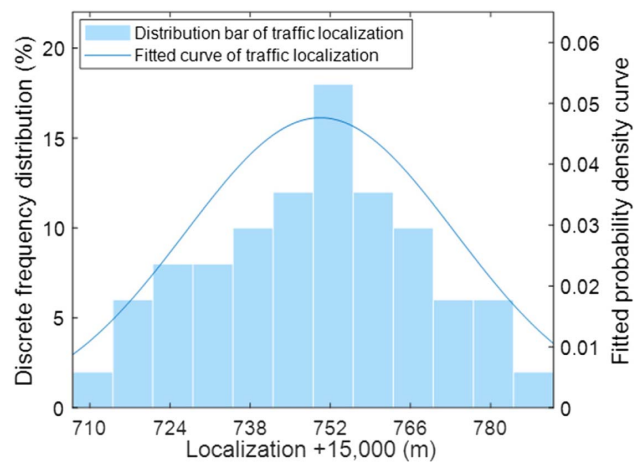

**Fig. 9.** Detection results of the urban link test. The blue bars are the distribution frequency of vibrations on Urban Link 1, and the blue line is the fitted probability density curve.

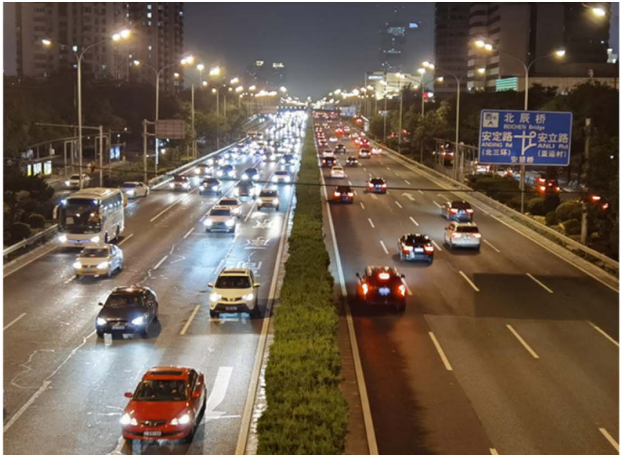

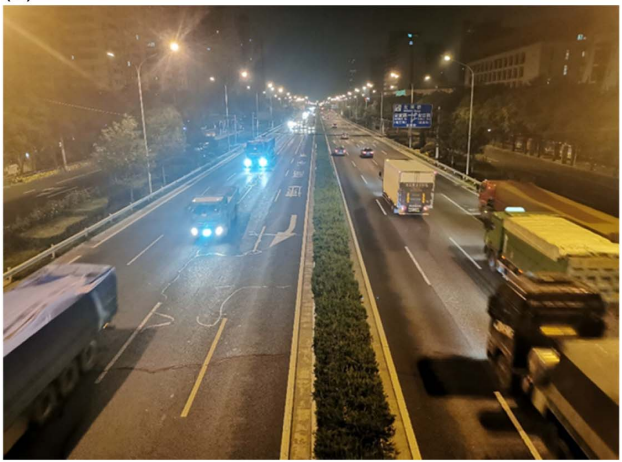

**Fig. 10.** Photos of the traffic condition at the target vibration location. (a) The typical traffic picture before 23:00. Vehicles on the road are mainly light-weight cars, which will cause low-amplitude high-frequency fluctuations. (b) The typical traffic picture after 24:00. Cargo lorries are permitted to enter the downtown city, which will produce high-amplitude low-frequency shockwaves.

**Funding.** National Natural Science Foundation of China (62171249, 61971259, 91836301).

**Acknowledgment.** We thank Professor Hongbo Sun of Tsinghua University and Dr. Pengbo Zhang of AutoNavi for fruitful discussions. We thank Jintao Yin of Beijing Gongjian Hengye Communication Technology Corporation Limited for arranging the urban fiber link.

**Disclosures.** The authors declare no conflicts of interest.

**Data Availability.** Data underlying the results presented in this paper are not publicly available at this time but may be obtained from the authors upon reasonable request.

†These authors contributed equally to this work.